\documentstyle[preprint,aps,psfig,tighten]{revtex}

\begin{document}

\draft

\title{Kinetic Partitioning Mechanism as a Unifying Theme in the
Folding of Biomolecules}

\author{D. Thirumalai, D.K. Klimov, and S.A. Woodson}
\address{Department of Chemistry and Biochemistry and \\
Institute for Physical Science and Technology\\
University of Maryland, College Park, Maryland 20742}

\maketitle

\begin{abstract}

We describe a unified approach to describe the kinetics of folding of
proteins and RNA.  The underlying conceptual basis for this framework
relies on the notion that biomolecules are topologically frustrated due to
the polymeric nature and due to the presence of conflicting energies. As a
result the free energy surface that has, in addition to the native basin
of attraction (NBA), several competing basins of attraction. A rough free
energy surface results in direct and indirect pathways to the NBA, i.e., a
kinetic partitioning mechanism (KPM). The KPM leads to a foldability
principle according to which fast folding sequences are characterized by
the folding transition temperature \(T_{F}\) being close to the collapse
transition temperature \(T_{\theta }\), at which a transition from the
random coil to the compact structure takes place. Biomolecules with
\(T_{\theta } \approx T_{F}\), such as small proteins and tRNAs, are
expected to fold rapidly with two-state kinetics. Estimates for the
multiple time scales in KPM are also given. We show that experiments on
proteins and RNA can be understood semi-quantitatively in terms of the
kinetic partitioning mechanism. 

\end{abstract}

\centerline{ \bf Introduction}
\nopagebreak
\vspace{7mm}

The pioneering experiments of Anfinsen and coworkers \cite{Anfinsen}
in the early sixties showed that the folding of proteins into a unique
structure with a well defined three-dimensional topology is a
self-assembly process, that is, the information needed to reach the
native conformation is encoded in the primary sequence
itself. Although these experiments were instrumental in demonstrating
the possibility that the native conformation of proteins corresponds
to the global free energy minimum they did not address the question of
how the native state is reached  in a biologically relevant time
scale, \(t_{B}\). The question of kinetic accessibility of the native
conformation came into focus when Levinthal argued \cite{Levin} that
the time for random search of all available conformations  far exceeds
\(t_{B}\). This seemed to imply that in order for proteins to reach
the native conformation on the time scale,  \(t_{B}\), there has to be
preferred pathways that will essentially limit the search of the
conformational space. The  Levinthal paradox, simplistic as it is, has
served as an intellectual impetus to understand how proteins find
their native conformations in a relatively short time. 

In the last several years through a combination of sophisticated
experiments
\cite{Englander,Bai,Udgaonkar,Roder,Rad,Matou,Fersht,Jones} 
and a study of minimal models of proteins 
\cite{Bryn89,Honey90,Shakh91,Leop,Chan94,Zwan92,Miller,Skol90,Skol91,Zwan95}
a new framework
for understanding folding kinetics has emerged. Recently several
reviews summarizing various aspects of the theoretical framework have
appeared \cite{Bryn95,Dill,Dill97}. 
In this article we describe a complementary but a different
perspective on how  biomolecules (both proteins and RNA) 
reach their native conformations
under folding conditions. The basic idea behind all theoretical
studies is that the
underlying topography of the free energy landscape of biomolecules is
rugged \cite{Bryn95,Dill97},  
consisting of many minima separated by barriers of varying
heights. It should be stressed that the  notion of a complicated free energy
surface (FES) has been invoked in a variety of contexts. The rugged
nature of potential energy surface  was introduced to understand slow
dynamics in structural 
glasses nearly thirty years ago \cite{Goldstein}. More recently, extensive
investigations have revealed that the hallmark of several classes of
disordered systems \cite{Mezard} is that the energy surface 
is complex, which in
turns leads to activated scaling and non-Arrhenius temperature
dependence of transport coefficients. Natural
proteins are unique in the sense that despite having a complicated FES
there is a dominant basin of attraction that is accessible on the time
scale \(t_{B}\). Therefore, the challenge is to understand how a
polypeptide chain explores the complex FES in order to reach the
global free energy minimum. A natural, but tautological, answer is that
biomolecular sequences have 
evolved to fold rapidly. The major contribution of theoretical
studies is to show how the rapid assembly of proteins and RNA takes place 
by examining the kinetic processes that are 
involved in the exploration of the complicated FES. In addition, 
spurred in part by the theoretical developments 
new experimental techniques have been elaborated to provide 
details (almost at the atomic level) 
of the events in the folding
process ranging from the tens of nanoseconds 
to submillisecond time scales. The joint efforts of the theoreticians and
experimental community are leading to rapid progress in our
ability to describe, in some instances quantitatively, the kinetics of
biomolecular self-assembly.  

In a recent article \cite{ThirumWood} we have suggested  that
concepts developed in the context of protein folding may be used to
understand the kinetics of RNA folding. 
The purpose of
this review is to describe a unified approach that leads to
uncover the global features
that are expected in the folding kinetics of proteins and RNA. We also provide
comparisons between theoretical predictions and experimental
measurements. Our goal is to point out that 
one can describe in some detail the
general kinetic principles of folding of biomolecules from the
statistical mechanics perspective. 
Furthermore, we will show that the expressions for the various time
constants can be estimated (at least, to within an order of magnitude) in
terms of  experimental parameters leading to validation of  
the theoretical concepts.

\vspace{7mm}
\centerline{ \bf  Topological Frustration and the Kinetic Partitioning
Mechanism}
\vspace{7mm}

Most of the qualitative features of the folding kinetics of
biomolecules can be understood by introducing the notion of
topological frustration. A crucial feature of proteins  is
that the primary sequence contains a certain fraction of monomers
that are hydrophobic. The fraction of hydrophobic 
residues in globular proteins  is in slight
excess of 0.5 \cite{White}. The linear density of the hydrophobic residues is
roughly uniform along the contour of the polypeptide chain which  means
that the hydrophobic residues are dispersed through out the chain. If
this were not the case, proteins would tend to aggregate.
A consequence 
of the uniform density of hydrophobic residues is that
on an any length scale \(l\), which is not equal to the size of the
chain, there is a propensity for the hydrophobic residues to form
tertiary contacts under folding conditions. The resulting structures,
formed by having contacts between proximal hydrophobic residues, would
in all likelihood be  incompatible with the unique global fold. The
incompatibility of the structures on local length scales with the
native conformation leads to topological frustration. 

It is important
to appreciate that topological frustration is an inherent
consequence of the polymeric nature of proteins 
(connectivity of residues) as well
as the presence of competing interactions (hydrophobic species, which 
prefer to form compact structures, and hydrophilic residues, which
are better accommodated by extended conformations). Thus, all proteins are
topologically  frustrated, with long chains being more so than smaller
ones. A direct consequence of topological frustration in
proteins is that the underlying topography of the free energy surface
is complex, consisting of many minima that are separated by a
distribution of barriers. We have recently shown \cite{ThirumWood}
that similar considerations apply to RNA as well. 

The nature of the low lying minima is easy to describe
qualitatively. On any length scale \(l\)  there are numerous ways of
constructing structures that are in conflict with global fold. Many,
perhaps most, of these structures would have high free energies and be
consequently unstable to thermal fluctuations. However, certain of
these structures are truly stable low energy minima that represent the
conformations that   can have many structural features  in common with
the native state. The energetic difference between these low energy
misfolded structures and the native state can be easily compensated by
the entropy associated with the misfolded structures. Thus, such
structures potentially act as kinetic traps in a typical folding
experiment and slow down the rate of protein folding. 

The fundamentals of the kinetic
partitioning mechanism (KPM) can be deciphered 
from the concept that there are low energy minima (in which the
proteins are misfolded) separated by free energy
barriers from the native state. 
Imagine an ensemble of denatured molecules under folding
conditions (achieved by diluting the concentration of the denaturant, 
for example) in search of the deepest basin of attraction in the rugged
free energy landscape. A fraction of the molecules \(\Phi \) would map
onto the native basin of attraction (NBA) directly and would reach it rapidly
without being trapped in other states. The remaining fraction would
inevitably be stuck in one or several of the low energy misfolded
structures, and only on a longer time scale reach the native state by
suitable activation processes. Thus, because of the topological
frustration that gives rise to a rugged free energy landscape, the
pool of denatured molecules partitions into fast folders and 
slow folders that reach 
the native state by indirect slow off-pathway processes. 

A schematic sketch of the KPM is shown in Fig. (1). The outline
of the kinetic scheme that emerges from Fig. (1) can be conveniently 
written as \\
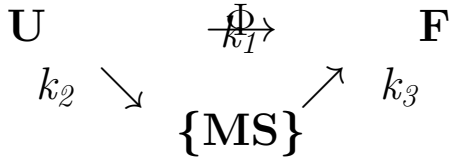
\begin{center}
{\Large
{\bf U } \hspace{14mm} $ \begin{array}{c}
\Phi \\[-7mm]
\longrightarrow \\[-7mm]
\normalsize{\em k_{1}}
\end{array}
$  \hspace{14mm} {\bf F}  \nopagebreak \\
$ \normalsize{\em k_{2}} \ \ \bf{\searrow }$  \hspace{17mm} $  \bf{
\nearrow } \ \  \normalsize{\em k_{3}}$ \nopagebreak  \\
\ \  {\bf \{MS\} } 
}
\end{center}
\noindent 
where { \bf U} refers to the  unfolded states, {\bf F } is the folded
state, and {\bf \{MS\} } denotes the collection  
of the low energy misfolded states. For
simplicity we have not indicated the rates for the backward processes in
above kinetic scheme.

The yield of  the fast process is given by \(\Phi \), the partition
factor.  Since  the shape and structure of the underlying FES
determine \(\Phi \), it becomes apparent that  \(\Phi \) depends not
only on the factors intrinsic to sequences but also on external
conditions such as pH, temperature and ionic strength. Furthermore,
\(\Phi \) can be easily altered by mutations so that a wild type
protein  with \(\Phi \approx 1\) can be made into a slow or moderate
folder. In the subsequent sections we will discuss in some detail the
dominant time scales that arise in the KPM. We will also show that
various experiments can be at least qualitatively understood in terms
of the kinetic partitioning mechanism.

\vspace{7mm}
\centerline{ \bf  Experimental Evidence for KPM}
\vspace{7mm}

There are several recent experimental studies  which support the basic
ideas of the kinetic partitioning mechanism
\cite{Rad,ThirumWood}. These experiments suggest that foldable
proteins can be divided into two classes;  first are fast folders
which reach the native conformation in a two  state kinetic process
without being trapped in any intermediates. Typically  fast folding
proteins are relatively small. From our theoretical perspective one
would conclude that the \(\sigma \) values (see Eq. (\ref{sigma})
below)  for these are small and the underlying energy landscape is
dominated by a single basin of attraction corresponding to the native
conformation. The time constant for reaching the native conformation
is of the order of a few milliseconds, which is consistent with the
theoretical prediction given in Eq. (\ref{tauNCNC}).  These proteins
have the partition factor, \(\Phi \), that is quite close to unity. 

The other class of proteins is the moderate folders which exhibit
multiple kinetic phases predicted by the KPM. The most detailed
verification of KPM has come in the study of the refolding of
lysozyme. In important early experiments Radford {\em et al}
\cite{Rad,Rad2} observed that the protection kinetics in hydrogen
exchange labeling experiments is well described by biphasic kinetics.
If we follow the interpretation suggested by Thirumalai and Guo
\cite{ThirumGuo} that the fast phase in these experiments describes
the mechanism of refolding to the native state by the
nucleation-collapse process, then the corresponding amplitude gives an
estimate of the partition factor. With this interpretation the
experiments of Radford {\em et al} \cite{Rad,Rad2} suggest that \(\Phi
\approx 0.25\) at \(T\,=\,20^{\circ}C\), pH\,=\,5.2  and at the
concentration of guanidinium chloride (GdmCl)  of \(0.54\,M\). 

More recently, Kiefhaber \cite{Kiefhaber1} 
has performed an ingenious experiment that
very clearly shows the full range of kinetic behavior predicted by
KPM. He used interrupted refolding experiments on hen egg lysozyme to
directly measure  the value of \(\Phi \). The refolding experiments are
initiated by diluting completely unfolded lysozyme into final folding
conditions (0.6\,M GdmCl, pH\,=\,5.2, and
\(T\,=\,20^{\circ}C\)). At various times the folding is interrupted by
transferring the solution to 5.3\,M GdmCl and pH\,=\,1.8. Apparently, under
these conditions the native lysozyme unfolds completely in about \(20
\, s\), while any partially formed misfolded structures unravel in a
few milliseconds. Thus, the amplitude of the slow unfolding process
gives the amount of the native structure when the folding process is
interrupted. By varying the time of interruption a history of the
kinetics of formation of the native state can be constructed. Kiefhaber
shows, by analyzing the folding kinetics at 0.6\,M GdmCl, that the
partition factor \(\Phi \) is \(0.14\), which implies that 14\% of the
initial population of denatured lysozyme 
reach the native state in \(50 \, ms\). The
majority of the population gets to the native conformation by indirect
processes that involves transitions out of the
kinetic traps. As shown below the time constant for this process
using Eq. (\ref{tauTSMM}) is about \(100 \,ms\), which is in
rough agreement with experimental estimate of 
\(420 \, ms\) \cite{Kiefhaber1}.

\vspace{7mm}
\centerline{ \bf  Dominant Time Scales in KPM}
\vspace{7mm}

The basic ideas leading to the KPM which are described above have
been substantiated using computer simulations of simplified models of
proteins \cite{ThirumWood,Mirny,Klim,Veitshans}. 
The dependence of the various parameters
characterizing KPM, namely, \(\Phi \), \(k_{1}, k_{2}, k_{3} \)
etc. on the properties  intrinsic to the sequence have been
identified \cite{Veitshans}. 
In particular, it has been shown in a series of papers
that for given external conditions 
the kinetic parameters of KPM are largely determined by
equilibrium temperatures that are  intrinsic to the 
protein sequence \cite{Klim,Veitshans}. It is now
known that for foldable sequences there are at least two equilibrium
temperatures that determine the "phases" of proteins
\cite{Bryn90,Cam93}. 
One is the high
temperature \(T_{\theta }\), at which the chain undergoes a transition
from a random coil conformation to a collapsed state; \(T_{\theta }\)
is very similar to the collapse transition temperature introduced by
Flory to describe the so called \(\theta \)-point in
homopolymers. Since there are several distinct energy scales 
 in proteins that discriminate
between the exponentially large number of compact conformations, the
chain undergoes a folding transition to the native state at
a temperature \(T_{F} \lesssim T_{\theta }\). The transition at
\(T_{F}\) is usually first order \cite{Cam93}, while the one at \(T_{\theta
}\) can be either first or second order depending upon a 
number of factors \cite{Cam97}, such as the relative strength of
effective two and three body interactions in the polypeptide chain. 
Both \(T_{\theta }\) and \(T_{F} \) are
experimentally measurable. The folding transition temperature, \(T_{F}
\), is usually associated with the midpoint of the temperature
dependence of denaturation plots. The collapse temperature is somewhat
harder to measure experimentally. Recently \(T_{\theta }\) has been
experimentally determined for few proteins \cite{Alex,Catan}. 

Using lattice models and Monte Carlo simulations and off-lattice
models and Langevin simulations we have established that folding times,
\(\tau _{F}\), correlate remarkably well with a single parameter
that is intrinsic to the sequence \cite{Klim,Veitshans,Cam93,Cam96}, namely
\begin{equation}
\sigma = (T_{\theta} - T_{F})/T_{\theta}. 
\label{sigma} 
\end{equation}
Therefore, \(\tau _{F}\) can be varied by altering \(T_{\theta }\) and
\(T _{F}\), both of which depend not only on the sequence
but also on the external conditions. 
It is clear that \(T_{F}\) depends precisely on the sequence and hence
can be altered by mutations. Surprisingly, \(T_{\theta }\) also
depends on the sequence \cite{Veitshans} 
(less sensitively than \(T_{F}\)). The reason
that \(T_{\theta }\) depends on the sequence and not just on the
sequence 
composition is that in addition to the hydrophobic interactions, 
due to the finite size of proteins the interfacial interactions 
between the surface residues and the solvent 
make a large contribution in determining the precise topology. 
The combination of hydrophobic and interfacial interactions 
determines \(T_{\theta }\) resulting in the collapse transition
temperature being sequence dependent.

The correlation between 
folding times and \(\sigma \) suggests that the rates of folding can
be altered by changing \(\sigma \) while leaving external conditions
fixed. Thus the wild type protein and a mutated one can have very
different folding rates depending upon \(\sigma \). This has also been
observed in RNA, in which the folding behavior can be
drastically altered by a single point mutation \cite{Jaeger,Emerick}. 
Using the concepts of polymer theory
one of us \cite{Thirum95} 
has shown that the time scales characteristic of KPM can
be established in terms of \(\sigma \) and other experimentally
controlled parameters. Remarkably enough the dominant time scales in the
folding process are once again controlled by \(\sigma \).

\vspace{7mm}
\centerline{ \bf  Fast processes and native conformation 
nucleation-collapse mechanism}
\vspace{7mm}

The fast processes, by which a certain fraction \(\Phi \) of the initial
ensemble of denatured molecules reaches the native state, have been
shown to occur via a native conformation nucleation-collapse (NCNC)
mechanism \cite{ThirumGuo,Abk1,Fer}. According to NCNC 
folding is initiated by the formation of native tertiary contacts. 
Once a critical
number of residues form tertiary native contacts, establishing an
overall  near-native
topology in the transition state, the polypeptide chain rapidly reaches
the native state. In this mechanism the processes of collapse and the
acquisition of the native state are almost synchronous
\cite{Thirum95}, and hence
would be nearly indistinguishable. 
The time scale for NCNC has
been argued to be \cite{Thirum95} 
\begin{equation}
\tau _{NCNC} \approx \frac{\eta a}{\gamma} \, f(\sigma ) \, N^{\omega
},
\label{tauNCNC}
\end{equation}
where \(\eta \) is the solvent viscosity, \(a\) is roughly the
persistent length of the protein, \(\gamma \) is the surface tension,
which tends to minimize the exposed surface area of the hydrophobic
species, and \(N\) is the number of amino acid residues in the
protein. The exponent \(\omega \) lies in the range \(3.8 \le \omega
\le 4.2\). The function \(f(\sigma )\) was originally shown to be
algebraic in \(\sigma \). Numerical studies indicate that \(f(\sigma
) \simeq \exp(\sigma / \sigma _{0})\) \cite{Klim,Cam96} 
provides a better fit to the
folding times. 

There are a number of remarks concerning NCNC and   \(\tau _{NCNC}\)
that are worth making: \\
\begin{enumerate}
\item If \(\sigma \) is small, which implies that
collapse and folding are almost synchronous, then \(\Phi \approx 1\)
and the folding time coincides with \(\tau _{NCNC}\). Typically this
is only expected for small proteins under optimal folding
conditions. Under these circumstances folding kinetics is expected to
display two-state behavior. Several experiments
suggest that small proteins exhibit the predicted
two state behavior \cite{Schindler,Krag}. 
\item  From a theoretical perspective sequences with small
\(\sigma \) are extremely well optimized so that the simultaneous
requirements of thermodynamic stability and the kinetic accessibility
of the native state can be achieved over a relatively large
temperature range. Numerical estimates \cite{Abk96} suggest that for these
sequences \(\tau _{F} \approx \tau _{NCNC}\) indeed scales
algebraically with \(N\) with \(\omega \approx 4\), confirming the
theoretical predictions \cite{Thirum95}. 
\item In a recent experiment Sch\"{o}nbrunner {\em et al} 
\cite{Kiefhaber} have suggested that for the 74-residue
all \(\beta \)-sheet forming protein tendamistat, collapse and folding are
essentially indistinguishable. Using the experimental parameters for
\(\eta \) and an estimate for \(\gamma \) and \(a\) the calculated
folding time according to Eq. (\ref{tauNCNC}) 
is about \(7 \, ms\), which is remarkably close to the
measured value of   \(10 \, ms\) \cite{Kiefhaber}. 
This estimate also suggests that experiments
in the submillisecond  regime are required to 
directly observe the nucleation dominated processes. 
\end{enumerate}

\vspace{7mm}
\centerline{ \bf  Three Stage Multipathway Mechanism}
\vspace{7mm}

In the case of moderate and slow folders it is likely that a large
fraction of initially denatured molecules does not reach the native
state by the NCNC mechanism described above. Such is the case in the
refolding of hen egg lysozyme, for example \cite{Kiefhaber1}.  The
partition factor \(\Phi \) under these circumstances is small. We now
describe the approach to the native state of the pool of molecules
that follow the indirect off-pathway processes. Extensive numerical
studies of lattice and off-lattice models show that the formation of
the native structure by indirect pathways can be conveniently
described by three stage multipathway kinetics  \cite{Rad2,Cam93}. A
brief description of each of the stages along with estimates of the
associated time scale is given below. 

\noindent 
(1) {\bf Non-specific collapse:} After the folding process is initiated the
polypeptide chain collapses into a relatively compact phase due to
the hydrophobic driving force. The kinetics in this stage is quite complex
and can perhaps be described by a distribution of time scales leading
to a stretched exponential behavior \cite{Cam93}. It should be
stressed that in
contrast to homopolymer collapse the initiation of collapse in
proteins is not completely random. The possible structures that are
seen in this phase could depend on loop formation probability,
dihedral angle transitions etc. It is likely that 
certain secondary structure elements such as helices, which form
rapidly, are already present at this stage. By a small generalization of the
arguments presented by de Gennes \cite{deGennes} 
the time scale for non-specific
collapse can be written as 
\begin{equation}
\tau _{c} \simeq \frac{\eta a}{\gamma } \, \biggl( \frac{T_{\theta }
-T_{F} }{ T_{\theta } }\biggr) ^{3} \, N^{2}
\end{equation}
An estimate for \(\tau _{c}\) can be made by taking \(\eta \approx
0.01 \, Poise\), \(a \approx (5\,-\,10)\,\AA\) and \(\gamma \approx
(40\,-\,60)\,cal/(\AA^{2}mol)\). With \(T_{F} \simeq T_{\theta }/2\),  
\(\tau _{c}\) is found to be between \((0.02\,-\,3)\,\mu s\) for
\(N=100\). This time scale roughly coincides with the time for forming small
number of contacts (see Eq. (\ref{tau})) between residues that are far
apart in the sequence space. 

\noindent 
(2) {\bf Kinetic Ordering: } In this stage the folding chain
discriminates between the exponentially large number of compact
conformations to form as many native contacts as possible which would
result in a lower free energy. In this phase free energy biases 
inherent in foldable sequences become operative and the chain
navigates to the CBA by a very cooperative motion. At the end of
this stage the chain reaches one of the low energy misfolded
structures which have many elements in common with the native
conformation. The search among these large number of compact
structures leading to the low energy misfolded conformations has been
argued to proceed by a reptation like process with the time constant
\cite{Thirum95} 
\begin{equation}
\tau _{ko} \approx \tau _{D} \, N^{\zeta }
\label{tau_stage2}
\end{equation}
with \(\zeta \approx 3\). The time constant \(\tau _{D}\) corresponds
to a local dihedral angle transition and is approximately \(10^{-8} \,
s\). Thus,  \(\tau _{ko} \approx 10 \, ms\) for \(N=100\). 

\noindent 
(3) {\bf All-or-None:} The last stage in the off-pathway processes 
involves activated transition from one of the many minimum energy
structures to the native state. This process necessarily involves
unraveling of the chain (at least partially) in order to break the
incorrect contacts and subsequently form the native
contacts. 
The partial unraveling of the chain in the process of
transition to the native state has been observed in 
numerical simulations of minimal models of proteins \cite{Chan94,Cam93}.  
It has been argued that the average free energy
barrier separating the misfolded structures and the native state
scales as \(\sqrt{N}k_{B}T_{F}\) \cite{Thirum95} under certain optimal folding
conditions,  so that the folding time for the slow
process is 
\begin{equation}
\tau _{F} \approx \tau _{0} \, e^{\sqrt{N}}
\label{tauTSMM}
\end{equation}
at \(T \simeq T_{F}\). Numerical simulations and more recently
experiments \cite{Hagen} suggest that \(\tau _{0} \simeq 10^{-6}\, s\) so that
\(\tau _{F}\) for \(N=100\) is \(0.1\,s\). 

Since the barrier height scales only sublinearly with \(N\) it is clear
that foldable sequences do not encounter the Levinthal paradox under
folding conditions even if they fold by indirect pathways. 
There are multiple pathways till the
second stage whereas only relatively few pathways connect the
misfolded structures and the native state. This is because, as
suggested elsewhere, the number of low energy compact structures only
scales as \(ln\,N\) \cite{CamSAW} .  
It is also clear that if
the molecules follow the three
stage kinetic approach to the native state then the transition states
occur closer to the native conformation.

\vspace{7mm}
\centerline{ \bf  Foldability Principle}
\vspace{7mm}

There are now numerous examples of proteins that reach 
the native state in few tens of
milliseconds under optimal
folding conditions \cite{Kiefhaber,Schindler,Krag}. 
The theoretical reasoning given above indicates
that under these conditions the value of \(\sigma \) for these
proteins is relatively small. These observations suggest a {\em foldability
principle} which can be stated as follows: {\em  A sequence is rapidly
foldable if \(T_{\theta } \approx T_{F}\)}. By foldability we mean
that both the kinetic accessibility and thermodynamic stability are
simultaneously satisfied. The foldability principle naturally applies
to small single domain proteins whose sequences can be optimized
relatively easily. 

It is, in fact, tempting to suggest that the
foldability principle, which express the kinetic accessibility
criterion in terms of properties intrinsic to the sequence, is a
quantitative realization of the consistency principle \cite{Go83} and the
principle of minimal frustration \cite{Bryn87}. Sometime ago Go
\cite{Go83} realized that 
spontaneous folding of proteins requires that the interactions
on short range, which are responsible for secondary structure formation,
be compatible with long range interactions, which confer global
topology. Here short and long refer to distances along the
sequence. Go also realized that it is not possible in nature to
produce an "ideal protein", in which there is a complete harmony
between long and short range interactions. More recently, Bryngelson
and Wolynes \cite{Bryn87} 
suggested, using the random energy model \cite{Der} as a paradigm for
protein folding, that the conflicts between various energy scales
should be minimized. 

If we take these principles into account we can
argue that the minimization of \(\sigma \) should be a natural
criterion for achieving a nearly ideal protein sequence. 
A heuristic argument leading to this 
conclusion goes as follows: The collapse transition temperature
\(T_{\theta }\) is primarily determined by a combination of 
the driving force that
tends to bury the hydrophobic residues in the core of the protein and
the forces that place the hydrophilic residues at the surface. The
free energy scale \(D\) determining \(T_{\theta }\) 
is obtained by a balance between the
hydrophobic interactions and the interfacial energies that tend 
to place hydrophilic residues at the surface to create a  
nearly compact structure. Thus, \(k_{B}T_{\theta } \approx
D\). We have recently argued \cite{Cam96} that \(T_{F}\) 
can be approximated as 
\begin{equation}
T_{F} \, \approx \, \frac{\Delta _{s}}{S_{NN}},
\end{equation}
where \(\Delta _{s} \) is roughly the stability gap and \(S_{NN}\) is
the entropy associated with the low energy non-native states. It is
reasonable to assume that \(S_{NN}\) also depends on \(D\). 
If the driving force is very large, then \(D/k_{B}T\,>>\,1\), and
the polypeptide chain will collapse into one of an 
exponentially large number
of conformations. Since \(\Delta _{s} \) is only weakly dependent on
\(N\) \cite{Thirum95} 
it follows that for large \(D\), \(T_{F} \approx 0\), which is
the homopolymer limit. In the opposite limit \(D/k_{B}T\, <<\, 1\)
there is not enough driving force for collapse and \(S_{NN}\) once
again grows with \(N\) (excluding logarithmic corrections) leading to
small \(T_{F}\). Sufficiently high value of \(T_{F}\) is obtained only when
\(S_{NN}\) is small or the number of low energy non-native structures
is not too big. Thus an optimum value of \(D\) is required so that
\(S_{NN}\) remains small. The existence of optimal \(D\) is also
consistent with the observation that in natural proteins the fraction
of hydrophobic residues is in slight excess of 0.5 \cite{White,CamSAW}. 
If we use the bound that \(T_{F} \lesssim
T_{\theta }\) then we see that  the optimal value for \(D\) results for
\(T_{F} \approx T_{\theta }\). An optimal value of \(D\) implies a proper
balance between long range and short range interactions so that
the hydrophobic interactions are in harmony with interfacial forces. Thus,
at least heuristically we can conclude that the consistency principle
\cite{Go83} 
and the principle of minimal frustration \cite{Bryn87} suggest that 
\(\sigma \)
should be small for optimally designed proteins. Since minimizing
\(\sigma \) seems probable only for small proteins it is tempting
to suggest that in nature bigger proteins are not optimized.

\vspace{7mm}
\centerline{ \bf  Early Events in Protein Folding}
\vspace{7mm}

The time scale estimate for \(\tau _{NCNC}\) for proteins that reach
the native state by  the nucleation-collapse folding process with
\(\Phi \approx 1\) is roughly  between \(0.1\,ms\) to few tens of
milliseconds depending on the length of the protein and external
conditions (see Eq.(\ref{tauNCNC})).  In addition, the time scale
(Eq. (\ref{tau_stage2})) for reaching  the low energy misfolded
structures by off-pathway processes  is also about few
milliseconds. It is, therefore, of interest to ask about processes
that take place on a submillisecond time scale.  Following the
pioneering work of Eaton and coworkers \cite{Jones,Eaton96}, there has
been an explosion of experimental papers
\cite{Chesick,Gruebele,Will,Yeh} probing protein folding events on
short time scales using a variety of techniques. In the original
experiments Jones {\em et al} \cite{Jones} used optical triggering to
refold cytochrome C in a  chemical denaturant. More recently
temperature jump \cite{Gruebele}, electron transfer \cite{Chesick},
and other novel mixing techniques \cite{Yeh} have been used to induce
and observe protein folding in submillisecond time scale. The major
conclusion of almost all these experiments is that significant
self-assembly of proteins   begins on time scales as small as a
microsecond. 

From a general perspective a question of some importance is whether
there is an upper limit for the rate of protein folding. In fact, in a
recent interesting article explaining the so called "new view" of
protein folding Dill and Chan \cite{Dill97} put down a "wish list" for
experimental studies of folding kinetics. One of the questions in  the
wish list is "What is the fastest speed  a protein can fold?". Hagen
{\em et. al.}  \cite{Hagen} have recently attempted to provide an
imaginative answer to this question by using the following
reasoning. They conducted an experiment to probe the time it takes for
two residues that are far apart in sequence space to form a
contact. Such a transient contact may either be native (i.e., the
contact is present in the final folded conformation) or
non-native. Using optical triggering to refold cytochrome C Hagen {\em
et. al.} \cite{Hagen} estimated that the diffusion controlled rate for
two sites separated by \(\sim 50-60\) residues to make a contact would
be  about \((100\, \mu s)^{-1}\). Using the loop formation probability
for stiff chains \cite{CamProx} they argued that the more probable
contacts between sites separated by 10 or 20 residues can occur in
about \(1\,\mu s\). Since the formation of a single tertiary contact
is the most elementary folding process (besides, say, the formation of
secondary structure like a helix) in the route towards the global fold
Hagen {\em et. al.}  \cite{Hagen} argued that the upper limit for
folding rate of a protein should be \(1\,(\mu s)^{-1}\). 

These experimental estimates that \(1\,\mu s\) should be an
important time scale in the initiation of certain events in the
folding process is consistent with theoretical estimates of the time
scales \(\tau (l)\) for diffusion
limited contact formation between two sites that are separated by
\(l\) residues. Guo and Thirumalai \cite{Guo} showed using scaling arguments
that \(\tau (l)\) can be estimated as 
\begin{equation}
\tau (l) \ \ \approx \ \ \frac{<R_{l}^{2}>}{P(l)D_{0}},
\end{equation}
where \(<R_{l}^{2}>\) is the spatial distance separating the two
sites, \(P(l)\) is the probability of loop formation \cite{CamProx}, 
and \(D_{0}\) is
the effective monomer diffusion constant. The distribution function
\(P(l)\) can be calculated by assuming that the backbone is stiff on
the scale of the order of persistent length and is given by \cite{CamProx}
\begin{equation}
P(l) \approx \Omega (N) \left\{ 
\begin{array}{ll}
0, \ \ l < l_{min}\\
\frac{ 1\,-\,\exp(-\frac{l}{l_{p}}) }{ l^{\theta _{3}}}, \ \ l \ge l_{min}
\end{array} 
\right.
\end{equation}
where \(l_{min}\) is length of the shortest loop possible, \(\theta
_{3}\) is an universal exponent whose value in three dimensions is
2.2, and \(l_{p}\) is an effective persistent length which measures
the stiffness of the polypeptide chain backbone. The normalization
factor \( \Omega (N)\) depends on the total number of residues. If we
use the Flory results for \(<R_{l}^{2}> \approx l_{p}^{2}l^{2\nu}\),
then the time constant \(\tau _{l}\) for \(l > l_{p}\) becomes
\begin{equation}
\tau _{l}\, \approx \,   \Omega (N) \frac{
l_{p}^{2}\,l^{2\nu+\theta_{3}} }{D_{0}}.
\label{tau}
\end{equation}
If we take the experimental estimates for \(D_{0} \simeq 10^{-6}\,
cm^{2}/s\) , \(l_{p} \approx 5\, \AA\), \(\nu \simeq 0.6\) then \(\tau
_{l}\) for \(l=10\) according to Eq. (\ref{tau}) turns out to be about
\(10\, \mu s\), where \(\Omega (N)\) is computed using \(l_{min}
\simeq 7\) and \(N \rightarrow \infty\). This theoretical estimate is
consistent with the experimental measurements given the inherent
uncertainties in \(D_{0}\) and \(l_{p}\). 

In retrospect it is not surprising that \(1\,\mu s\) or so turns out
to be an important time scale in the early processes of protein
folding. This was already realized based on theoretical arguments
given in the context of the refolding of bovine pancreatic trypsin
inhibitor \cite{CamProx,ThirumLecture}.  However, an understanding of
how the formation of these contacts leads to  further self-assembly of
proteins is still lacking.  This requires experiments that can
directly observe correlated events. It is only through such
experiments that a molecular basis for the nucleation-collapse
mechanism can be provided. 

Experimental \cite{Hagen}  and theoretical \cite{Guo}  studies clearly
suggest that transient long range contacts occur on some microsecond
time scale. If these contacts are non-native and stable then the
polypeptide chain will subsequently 
collapse into a misfolded structure on a short
(much less than about milliseconds) time scale. In proteins with small
\(\sigma \) the initial stable contacts are expected to be 
native, and once
a sufficient number of such contacts 
forms then a rapid transition to the native state
takes place presumably via the nucleation-collapse mechanism. 
As suggested above experiments that can observe correlation between
multiple contacts will be required to further elucidate the nature of
the nucleation-collapse mechanism. 

\vspace{7mm}

\centerline{ \bf  Folding of RNA  }
\nopagebreak 
\vspace{7mm}

The folding pathways of large RNA are now beginning to be probed
experimentally. In the last  year the existence of possible
connections between the folding of RNA and proteins  was pointed out
\cite{ThirumWood,Zarr}. From the perspective of the rough FES it is
natural to expect some common qualitative elements  for folding
kinetics of proteins and RNA \cite{ThirumWood}.  The usual arguments
like the incompatibility of times for random search of all
conformations and biological folding times apply equally well to
nucleotide sequences.  For RNA  it is necessary to form correct
secondary structures, namely Watson-Crick pairs between complementary
sequences. The correctly formed secondary structures  assemble to
achieve the correct three-dimensional organization of the structural
elements. Although we expect certain common trends in the folding of
RNA and proteins there are also fundamental differences
\cite{ThirumWood,Zarr}  that  have not been explored. One major
difference is that collapse of RNA typically requires binding of
divalent ions \cite{Gross,Pyle}. 

As for proteins, it is found that certain RNA sequences fold rapidly
without being trapped in misfolded states. These include tRNAs
\cite{Crothers} as well as certain small group I self-splicing introns
\cite{Krag}  for which \(\Phi \) under in vitro conditions appears to
be close to unity. The folding time for tRNA was estimated to be about
\(0.1\,-\,1\,s\) \cite{Crothers}  suggesting that perhaps folding
occurs via a nucleation-collapse mechanism. 

In a recent article \cite{ThirumWood} we have begun to analyze in
quantitative terms the folding kinetics of {\em Tetrahymena} ribozyme
in terms of the kinetic partitioning mechanism. The availability of
considerable structural information makes the {\em Tetrahymena}
ribozyme an attractive model for studying the folding of large RNAs
\cite{Cech}. The arguments given for topological frustration suggest
that the low energy misfolded structures become more prominent for
larger length chains resulting  in a smaller value of \(\Phi \)
\cite{ThirumWood}. 

These expectations are borne out in the quantitative analysis of the
experiments on the refolding of precursor RNA containing the {\em
Tetrahymena} ribozyme. The experiments of Emerick and Woodson
\cite{Emerick1994,Emerick1996} showed using self-splicing kinetics and
gel electrophoresis that a population containing a mixture of active
and inactive conformers is in slow exchange at
\(T=30^{\circ}C\). Majority of the population (\( > \, 70\%\)) of the
wild-type {\em Tetrahymena} precursor RNA appears to be misfolded
after transcription at \(T=30^{\circ}C\). If the RNA is heated to
\(T=75^{\circ}C\) and annealed to \(T=30^{\circ}C\) the percentage of
inactive molecules decreases to about \(20\,-\,30\,\%\). Thus the
inactive conformations (presumably misfolded) can be made to reach the
native state by an annealing process. 

The experimental findings of Emerick and Woodson
\cite{Emerick1994,Emerick1996}  confirm the basic
picture of folding predicted by KPM. According to KPM for larger RNAs
one expects the chain to be trapped in one of the low energy
structures. The relatively slow folding is a consequence of escape
from these traps by an activated process. The partition factor at
\(T=30^{\circ}C\) is small (\(\Phi \approx 0.2\)) which implies that
most of the molecules reach the native conformation by off-pathway
processes. 

Another theoretical prediction made in the context of proteins that
appears to be consistent with experiments on RNA is the activation 
energy separating the misfolded states and the active folded
conformation. Based on the temperature dependence of the conversion of
the inactive to active form the barrier height was estimated to be
\((10\,-\,15)\,kcal/mol\) \cite{Emerick1994,Emerick1996}. 
According to the theoretical arguments such
barriers are expected to scale as 
\(\sqrt{N}k_{B}T\) \cite{Thirum95}  which for {\em Tetrahymena} precursor RNA
(\(N=650\)) turns out to be \(15\,kcal/mol\). 
The good agreement between theoretical estimate and 
experiments suggests that typical free 
energy barriers in biomolecules are small.

\vspace{7mm}
\centerline{ \bf  Conclusions}
\vspace{7mm}

It is gratifying that certain general principles of folding kinetics
of proteins and RNA can be deciphered from simple considerations
\cite{ThirumWood}. Due  to their polymeric nature and the presence of
conflicting energy scales, biomolecules are intrinsically
topologically frustrated. As a consequence the free energy surface is
complex and contains not only the native basin of attraction (NBA),
but also competing basins  of attraction (CBA) as  well. The basic
features of the kinetic partitioning mechanism  (KPM) naturally emerge
from this idea.  The concepts outlined  in this article should be
viewed as a tentative unified proposal to conveniently classify the
possible scenarios that can arise in the complex self-assembly of
proteins and RNA. It is possible to extend these concepts to make
testable predictions for the folding of specific proteins
\cite{CamProx} and RNA, assuming  more detailed models that account
for solvent conditions and other aspects that are left out in the
simplified description. It is nonetheless clear that our understanding
of the folding kinetics will continue to grow rapidly through an
interplay between theoretical ideas and experimental advances.

A unified description of the folding process of proteins and RNA is
expected to advance the study of RNA folding. For example,  it is
logical to suggest that fast processes in RNA could well determine the
extent to which misfolded structures are going to slow down the
folding process. If the similarities between the nature of folding of
proteins and RNA are further pursued then it would imply that the
organization of the folded structure of RNA also involves parallel
pathways \cite{Pan}.  The recent semi-quantitative analysis
\cite{ThirumWood} of the experiments of Emerick and Woodson
\cite{Emerick1994} strongly suggests that folding of large RNAs does
take place by multiple parallel pathways. Additional  experiments on
faster time scale are needed to further elucidate the nature of these
pathways. 

Finally, the KPM also points to the need for chaperonin assisted
folding of proteins and RNAs \cite{ThirumWood}.  The arguments based
on KPM would suggest that only when the partition factor is small
(less than \(10\,\%\)) does one require the chaperonin machinery
\cite{ThirumWood,Todd}.  Typically, as suggested here and elsewhere
\cite{ThirumWood},  this happens only for large proteins. For these,
\(N\) is sufficiently large so that the folding time given by
Eq. (\ref{tauTSMM}) not only exceeds \(t_{B}\) but starts to become
comparable with the time scale for aggregation processes. Under these
circumstances the chaperones are predicted to rescue the misfolded
structures by a process referred to as iterative annealing mechanism
\cite{Todd,ChanDill1996}.  A similar reasoning would suggest that for
large RNA as well there must exist RNA chaperones \cite{ThirumWood}
which presumably function in a manner similar to GroEL and GroES. The
RNA chaperones have not yet identified, although certain non-binding
RNA proteins have been shown to enhance the rate of RNA-catalyzed
reactions in vitro  \cite{Tsuchihashi,Coetzee}. 

\acknowledgments

We would like to thank Oksana Klimova for preparing the figure. 
This work was supported in part by grants from the National Science
Foundation, Air Force Office of Scientific Research, National
Institutes of Health, American Cancer Society, Pew Charitable Trust,
and Camille and Henry Dreyfus Foundation.

\newpage
\centerline{FIGURE CAPTION}
\vspace{7mm}

Fig. 1.  A pictorial representation of the kinetic partition
mechanism.  The unfolded structures collapse rapidly (in, perhaps,
microsecond time  scales). These structures contain almost all of the
secondary structures. These structural elements are shown as blocks
labeled as A - G.  Some of these blocks show helices, which are
expected to form in submicrosecond time scales, while other blocks
show beta-strands, which form in about ten microseconds. The
subsequent packing of these secondary structural elements results in a
fraction of the population \(\Phi \) going directly to the native
state via the native conformation nucleation-collapse mechanism. An
example of a transition state obtained along this pathway is displayed
as an expanded version of the native conformation. This structure
contains all native-like contacts and the lack of native contacts
between  blocks A and G and B and F makes this structure somewhat
larger than the native state. The remaining fraction of the molecules,
\(1 - \Phi \), gets trapped in misfolded structures, an example of
which is shown in the upper right corner.  In this case helices A and
G have incorrect orientation and non-native contact between B and E
has been formed. The activated transitions from the misfolded
structures to the native state involve partial unraveling of the
polypeptide chain to break the incorrect contacts and establish native
contacts. In this highly simplified representation hydrophobic
portions of sequence are shown in blue and the hydrophilic are given
in red.


\begin{references}

\bibitem{Anfinsen} Anfinsen, C.A. {\em Science } {\bf 1973}, {\em 181,
223}.

\bibitem{Levin} Levinthal, C. In {\em Mossbauer spectroscopy in
biological systems}; Debrunner, P., Tsibris, J.C.M., M\"{u}nck, E.
Eds.; University of Illinois Press: Urbana, 1969. 

\bibitem{Englander} Englander, S.W.; Poulsen, A. {\em Biopolymers}
{\bf 1969}, {\em 7, 329}. 


\bibitem{Bai} Bai, Y.; Englander, S.W.  {\em Proteins
Struct. Funct. Genet.} {\bf 1996}, {\em 24, 145}. 

\bibitem{Udgaonkar} Udgaonkar, J.B.; Baldwin, R.L. {\em Nature} {\bf
1988}, {\em 335, 694}.

\bibitem{Roder} Roder, H.; El\"{o}ve, G.A.; Englander, S.W. {\em Nature}
{\bf 1988}, {\em 335, 700}. 


\bibitem{Rad} Radford, S.E.; Dobson, C.M.   
{\em Phil. Trans. Roy. Soc. Lond. B} {\bf 1995}, {\em 348, 17}. 

\bibitem{Matou} Matouschek, A.; Kellis, J.T., Jr.; Serrano, L.;
Bycroft, M.; Fersht, A.R. {\em Nature} {\bf 1990}, {\em 346, 440}. 

\bibitem{Fersht}  Fersht, A.R. {\em Curr. Opin. Struct. Biol.} {\bf
1995}, {\em 5, 79}. 


\bibitem{Jones} Jones, C.M.; Henry, E.R.; Hu, Y.; Chan, C.K.; Luck,
S.D.; Bhuyan, A.; Roder, H.; Hofrichter, J.; Eaton, W.A. 
{\em Proc. Natl. Acad. Sci. USA } {\bf 1993}, {\em 90, 11860}. 

\bibitem{Bryn89} Bryngelson, J.D.; Wolynes, P.G. 
{\em J. Phys. Chem. } {\bf 1989}, {\em 93, 6902} .


\bibitem{Honey90} Honeycutt, J.D.; Thirumalai, D. {\em Proc. Natl. 
Acad. Sci. USA} {\bf 1990}, {\em  87, 3526}.

\bibitem{Shakh91} Shakhnovich, E.; Farztdinov, G.; Gutin, A.M.; 
Karplus, M. {\em Phys. Rev. Lett.} {\bf 1991}, {\em 67, 1665}.

\bibitem{Leop} Leopold, P.E.; Montal, M.; Onuchic, J.N. 
{\em Proc. Natl. Acad. Sci. USA} {\bf 1992}, {\em  89, 8721}.

\bibitem{Chan94} Chan, H.S.; Dill, K.A.  {\em J. Chem. Phys.} {\bf
1994}, {\em 100, 9238}. 

\bibitem{Zwan92} Zwanzig, R.; Szabo, A.; Bagchi, B. 
{\em Proc. Natl. Acad. Sci. USA} {\bf 1992}, {\em  89, 20}.

\bibitem{Miller} Miller, R.; Danko, C.A.; Fasolka, M.J.; Balazs, A.C.;
Chan, H.S.; Dill, K.A. {\em  J. Chem. Phys.} {\bf 1992}, {\em 96, 768}. 


\bibitem{Skol90} Skolnick, J.; Kolinski, A.  {\em Science} {\bf 1990},
{\em 250, 1121}. 


\bibitem{Skol91} Skolnick, J.; Kolinski, A. 
{\em J. Mol. Biol.} {\bf 1991}, {\em  221, 499}.


\bibitem{Zwan95} Zwanzig, R. {\em Proc. Natl. Acad. Sci. USA} {\bf
1995}, {\em  92, 9801}.

\bibitem{Bryn95} Bryngelson, J.D.; Onuchic, J.N.; Socci, N.D.; Wolynes,
P.G.  {\em Proteins Struct. Funct. Genet.} {\bf 1995}, {\em  21, 167}. 


\bibitem{Dill} Dill, K.A.; Bromberg, S.; Yue, K.; Fiebig, K.M.; Yee, D.P.;
Thomas, P.D.; Chan, H.S. {\em  Protein Sci.} {\bf 1995}, {\em 4,
561}. 

\bibitem{Dill97} Dill, K.A.; Chan, H.S. {\em Nature Struct. Biol.}
{\bf 1997}, {\em 4, 10}. 

\bibitem{Goldstein} Goldstein, M. {\em J. Chem. Phys.} {\bf 1969}, {\em
51, 3328}. 

\bibitem{Mezard} {\em Spin Glass Theory and Beyond}; Mezard, M.;
Parisi, G.; Virasaro, M.A.; World Scientific: Singapore, 1987. 

\bibitem{ThirumWood} Thirumalai, D.; Woodson, S.A. {\em
Acc. Chem. Res.} {\bf 1996}, {\em 29, 433}. 

\bibitem{White} White, S.H.; Jacobs, R.E. {\em J. Mol. Evol.}
{\bf 1993}, {\em  36, 79}. 

\bibitem{Rad2} Radford, S.E.;  Dobson, C.M; Evan, P.A. {\em Nature}
{\bf 1992}, {\em 358, 302}. 


\bibitem{ThirumGuo} Thirumalai, D.; Guo, Z.   
{\em Biopolymers} {\bf 1995}, {\em 35, 137}. 

\bibitem{Kiefhaber1} Kiefhaber, T. {\em Proc. Natl. Acad. Sci. USA} 
{\bf 1995}, {\em 92, 9029}. 
 
\bibitem{Mirny} Mirny, L.A.; Abkevich, V.;   Shakhnovich,
E.I. {\em Folding \& Design} {\bf 1996}, {\em 1, 103}. 

\bibitem{Klim} Klimov, D.K.; Thirumalai, D. 
{\em Proteins Struct. Funct. Genet.} {\bf 1996}, {\em 26, 411}. 

\bibitem{Veitshans} Veitshans, T.; Klimov, D.K.; Thirumalai, D. 
{\em Folding \& Design} {\bf 1997}, {\em 2, 1}. 

\bibitem{Bryn90} Bryngelson, J.D.; Wolynes, P.G. {\em Biopolymers}
{\bf 1990}, {\em 30, 177}. 

\bibitem{Cam93} Camacho, C.J.; Thirumalai, D. 
{\em Proc. Natl. Acad. Sci. USA } {\bf 1993}, {\em 90, 6369}.

\bibitem{Cam97} Camacho, C.J.; Schanke, T. {\em Europhys. Lett.} (in
press). 

\bibitem{Alex} Alexander, P.; Fahnestock, S.; Lee, T.; Orban, J.; Bryan,
P. {\em Biochemistry} {\bf 1992}, {\em 31, 3597}. 

\bibitem{Catan} Catanzano, F.; Giancola, C.; Barone, G. {\em
Biochemistry} {\bf 1996}, {\em 35, 13378}. 

\bibitem{Cam96} Camacho, C.J.;  Thirumalai, D. 
{\em Europys. Lett.} {\bf 1996}, {\em 35, 627}. 

\bibitem{Jaeger} Jaeger, L.; Westhoff, E.; Michel, F. {\em
J. Mol. Biol.} {\bf 1993}, {\em 234, 331}. 

\bibitem{Emerick} Emerick, V.L.; Woodson, S.A. {\em Biochemistry} {\bf
1993}, {\em 32, 14062}. 

\bibitem{Thirum95} Thirumalai, D. {\em  J. Physique (Paris)
I} {\bf 1995}, {\em  5. 1457}.

\bibitem{Abk1} Abkevich, V.I.; Gutin, A.M.; Shakhnovich, E. 
{\em Biochemistry } {\bf 1994}, {\em  33, 10026}.

\bibitem{Fer} Fersht, A.R. 
{\em Proc. Natl. Acad. Sci. USA} {\bf 1995}, {\em  92, 10869}; 
Sosnick, T.R.; Englander, S.W. {\em Proteins Struct. Funct. Gen.} {\bf
1996}, {\em 24, 413}. 


\bibitem{Abk96} Abkevich, V.I.; Gutin, A.M.; Shakhnovich, E.I. {\em
Phys Rev. Lett.} {\bf 1996}, {\em 77, 5433 }. 

\bibitem{Kiefhaber} Sch\"{o}nbrunner, N.; Koller, K.-P.; Kiefhaber,
T. {\em J. Mol. Biol.} (submitted, 1996). 


\bibitem{Schindler} Schindler, T.; Herrler, M.,; Marahiel, M.A;
Schmid, F.X. {\em Nature Struct. Biol.} {\bf 1995}, {\em 2, 663}. 

\bibitem{Krag} Kragelund, B.B.; Robinson, C.V.; Kundsen, J.; Dobson,
C.M.; Poulsen, F.M. {\em Biochemistry} {\bf 1995}, {\em 34, 7217}. 

\bibitem{deGennes} de Gennes, P.G. {\em J. Phys Lett.} {\bf
1985}, {\em 46, L639}. 


\bibitem{Hagen} Hagen, S.J.; Hofrichter, J.; Szabo, A.; Eaton, W.A. {\em
Proc. Natl. Acad. Sci. USA} {\bf 1996}, {\em 93, 11615}. 

\bibitem{CamSAW} Camacho, C.J.; Thirumalai, D. {\em Phys. Rev. Lett.}
{\bf 1993}, {\em 71, 2505}. 

\bibitem{Go83} Go, N. 
{\em Ann. Rev. Biophys. Bioeng. } {\bf 1983}, {\em 12, 183}. 


\bibitem{Bryn87} Bryngelson, J.D.; Wolynes, P.G. {\em
Proc. Natl. Acad. Sci. USA} {\bf 1987}, {\em 84, 7524}. 

\bibitem{Der} Derrida, B. {\em Phys. Rev. Lett.} {\bf 1980}, {\em  45,
79}.

\bibitem{Eaton96} Eaton, W.A.; Thompson, P.A.; Chan, C.K.; Hagen, S.J.;
Hofrichter, J. {\em J. Structure} {\bf 1996}, {\em 4, 1133}. 

\bibitem{Chesick} Pascher, T.; Chesick, J.P.; Winkler, J.R.; Gray,
H.B.    {\em Science} {\bf 1996}, {\em 271,
1558}. 

\bibitem{Gruebele} Ballew, R.M.; Sabelko, J.;  Gruebele, M.
{\em Proc. Natl. Acad. Sci. USA
} {\bf 1996}, {\em 93, 5759}. 

\bibitem{Will} Williams, K.; Cansgrove, T.P.; Gilmanshin, R.; Fang, K.S.;
Callander, R.H.; Woodroff, W.H.; Dyer, R.B. {\em Biochemistry} {\bf 
1996}, {\em 35, 691}. 

\bibitem{Yeh} Yeh, S.R.; Takahashi, S.; Fan, B.; Roussean, D.L. {\em
Nat. Struct. Biol.} {\bf 1997}, {\em 4, 51}. 

\bibitem{CamProx} Camacho, C.J.; Thirumalai, D. 
{\em Proc. Natl. Acad. Sci. USA} {\bf 1995}, {\em 92, 1277}. 


\bibitem{Guo} Guo, Z.; Thirumalai, D. {\em Biopolymers} {\bf 1995},
{\em  36, 83}.

\bibitem{ThirumLecture} Thirumalai, D. Lectures in Statistical
Mechanics. Protein Structure, NATO Workshop, Corsica, 1993. 

\bibitem{Zarr} Zarrinkar, P.P.; Williamson, J.R. {\em Nat. Struct. Biol.}
{\bf 1996}, {\em 3, 432}. 

\bibitem{Gross} Lindahl, T.; Adams, A.; Fresco, J.R. {\em Proc. 
Natl. Acad. Sci. USA} {\bf 1966}, {\em 35 1941};  
Grosshans, C.A.; Cech, T.R. {\em Biochemistry} {\bf
1984}, {\em 28, 6888}. 

\bibitem{Pyle} Pyle, A.M. {\ Science} {\bf 1993}, {\em 261, 709}. 

\bibitem{Crothers} Crothers, D.M.; Cole, P.E.; Hilbers, C.W.; Schulman,
R.G. {\em J. Mol. Biol.} {\bf 1974}, {\em 87, 63}. 

\bibitem{Cech} Cech, T.R. In {\em The RNA World} Gesteland, R.F.;
Atkins, J.F. Eds. ; Cold Spring Harbor Press: New York, 1993. 

\bibitem{Emerick1994} Emerick, V.L.; Woodson, S.A. {\em
Proc. Natl. Acad. Sci. USA } {\bf 1994}, {\em 91, 9675}. 

\bibitem{Emerick1996} Emerick, V.L.; Pan, J.;  Woodson, S.A.
{\em Biochemistry} {\bf 1996}, {\em 35, 13469}. 

\bibitem{Pan}  Pan, J.; Woodson, S.A.; Thirumalai, D. (to be
published). 

\bibitem{Todd} Todd, M.J.; Lorimer, G.H.; Thirumalai, D. 
{\em Proc. Natl. Acad. Sci. USA } {\bf 1996}, {\em 93, 4030}. 

\bibitem{ChanDill1996} Chan, H.S.; Dill, K.A. {\em Proteins
Struct. Funct. Gen.} {\bf 1996}, {\em 24, 345}. 

\bibitem{Tsuchihashi} Tsuchihashi, Z.; Khosla, M.; Herschlag, D. {\em
Science } {\bf 1993}, {\em 262, 99}. 

\bibitem{Coetzee} Coetzee, T.; Herschlag, D.; Belfort, M. 
{\em Genes Dev.} {\bf 1994}, {\em 8, 1575}. 




\end{references}
\end{document}